\documentclass[doublecol,reprint]{epl2} 
\usepackage{amsmath,amssymb}
\usepackage{verbatim}
\usepackage{graphicx}
\usepackage{color}
\usepackage[colorlinks,bookmarks=false,citecolor=blue,
linkcolor=red,urlcolor=blue]{hyperref}
\usepackage{times}
\usepackage{soul}

\def \beq{\begin{equation}}
\def \eeq{\end{equation}}
\def \barray{\begin{eqnarray}}
\def \earray{\end{eqnarray}}

\begin{document}

\author{Mohammad Walid AlMasri}
\institute{
Department of Physics, Ko\c{c} University, Rumelifeneri Yolu, 34450 Sar\i yer, 
Istanbul, Turkey.  
}

\date{\today}

\title{$SU(4)$ description of bilayer skyrmion-antiskyrmion pairs}  

\abstract{
 The antiferromagnetic coupling and entanglement  between skyrmion lattices are treated in magnetic bilayer systems. We first formulate the problem of large bilayer skyrmions using $\mathbb{CP}^{1}\otimes \mathbb{CP}^{1}$ theory. We have considered bilayer skyrmions under the presence of Dzyaloshinskii-Moriya (DMI) and Zeeman interactions confined in a two-dimensional chiral magnet such as Fe$_{0.5}$Co$_{0.5}$Si. We parametrize  bilayer skyrmions using $SU(4)$ representation, and represent each skyrmion and antiskyrmion using Schmidt decomposition. The reduced density matrices for skyrmion and antiskyrmion has been calculated. The conditions for maximal, partial entanglement and separable bilayer skyrmions are presented. Our results can be used for generating entanglement in systems with large number of spins.}

\maketitle

\section{Introduction}
Magnetic Skyrmions are microscopic topological defects in  spin textures that are characterized by the charge \cite{Nagaosa},  
\begin{equation}
Q=\frac{1}{4\pi}\int d^{2}{\bf r} 
\;  {\bf n} . \; (\frac{\partial {\bf n}}{\partial x}\times \frac{\partial {\bf n}}{\partial y})=\pm 1 . 
\end{equation} 
In mathematical literature, $Q$  is a  topologically invariant quantity known as  Pontryagin number. It counts how many times ${\bf n}({\bf r})={\bf n}(x,y)$ wraps the unit sphere \cite{Nash}. Skyrmions were first introduced by  Skyrme \cite{Skyrme1} to explain hadrons in  nuclei. Interestingly it has  turned out to be relevant in other condensed matter systems such as chiral magnets \cite{Muhlbauer}. Theoretically, magentic skyrmions were introduced and investigated by Bogdanov and his collaborators  in \cite{Bogdanov, Bogdanov1} . 
Skyrmions can be driven by charge or spin currents  in confined geometries \cite{torque}. In general, skyrmions are subject to skyrmion Hall effect (SkHE) caused by Magnus force.  SkHE was predicted theoretically in \cite{SkHE} and has been observed experimentally \cite{SkHE1}. Magnus force is the force acting transverse to the skyrmion velocity in the medium and can be interpreted as a manifestation of the real-space Berry phase \cite{Sonin,Freimuth}. 

SkHE is a detrimental effect since the skyrmions experience it will deviate from going in straight path. As a result, moving skyrmions can be damaged or even  destroyed  at the edges of  thin film sample. One way  of suppressing SkHE is to consider two perpendicular chiral thin films strongly coupled via  antiferromagnetic (AFM) exchange coupling.  It is expected that  when skyrmion lattice is formed at the bottom thin film , simultaneously another skyrmion lattice is  created at the top thin film  with opposite topological charge. In this case, the SkHE is vanished since the Magnus force acting on the  top  skyrmion (antiskyrmion) is equal to the Magnus force that acts on the bottom antiskyrmion (skyrmion) with opposite sign leaving us with zero net force.
Analogous scheme was proposed to suppress SkHE in nanoscale N{\'e}el skyrmions by considering two perpendicular ferromagnetic films separated by an insulator with heavy metal underneath the second ferromagnetic film   \cite{Ezawa}. \\  
Quantum signatures for large skyrmions  can emerge at  the phase boundary between skyrmion crystal phase (SkX) and ferromagnetic phase at zero temperature like skyrmions in  Fe$_{0.5}$Co$_{0.5}$Si. During this phase transition a quantum liquid phase is expected to emerge \cite{Takashima}. In this case, the classical LLG \cite{Gilbert} and Thiele equation \cite{Thiele} break down due to quantum fluctuations. The full quantum theory of bilayer skyrmions is out of the scope of this work and it can be recovered under some circumstances. As an example, for sufficiently weak antiferromagnetic exchange coupling  between thin films, bilayer skyrmion (antiferromagnetically coupled skyrmion-antiskyrmion pair) can be seen as two separate skyrmions and the quantum dynamics is already known for a single large skyrmion \cite{Takashima}. In this work, we  give a detailed theory of large  bilayer skyrmions ( with sizes at order of 100 nm )  using  HDMZ (Heisenberg exchange $+$ Dzyaloshinskii-Moriya interaction $+$ Zeeman interaction) model. We study the problem of entanglement in large bilayer skyrmions  from general perspective using our developed continuum theory of bilayer skyrmions and $SU(4)$ representation. In final section, we study the geometry of quantum states in bilayer skyrmions. 
\section{The $\mathbb{CP}^{1}\otimes \mathbb{CP}^{1}$-Theory of Large Bilayer Skyrmion } \label{continuum}

	We consider two  thin films of chiral magnets  separated by an insulating spacer with antiferromagnetic coupling between chiral  films. We assumed each film to host Bloch skyrmions under certain ranges of temperature and external magnetic field determined by the film parameters.  Skyrmions in the first thin-film are equal in size with skyrmions in the second thin-film but with opposite topological charge. 
	For our model to hold, we assume temperatures lower than the magnon gap and skyrmion with large radius \cite{Takashima}. Fortunately,   skyrmions in Fe$_{0.5}$Co$_{0.5}$Si support these assumptions \cite{Feco}.
We present a detailed theory of bilayer skyrmions written with respect to  $\mathbb{CP}^{1}\otimes \mathbb{CP}^{1}$-theory. The HDMZ Hamiltonian density for each chiral magnet layer is 
\begin{equation}\label{effective}
\mathcal{H}_{i}= \frac{J}{2} \;   (\partial_{\mu}{\bf n}^{i})\cdot (\partial_{\mu}{\bf n }^{i})+ D \; {\bf n}^{i}\cdot(\nabla \times{\bf n}^{i}) - {\bf B}\cdot{\bf n}^{i}, 
\end{equation}
We adopted Einstein summation notation for repeated indices $\partial_{\mu}{\bf n}\cdot \partial_{\mu}{\bf n}\equiv \Sigma_{\mu} \partial_{\mu}{\bf n}\cdot \partial_{\mu}{\bf n}$. Since we are interested in two-dimensional thin films $\mu=x,y$. The index $i=S,A$ labels the skyrmion and anti-skyrmion respectively and 
${\bf n^{i}}=(\sin\theta^{i}\cos\phi^{i},  \sin\theta^{i}\sin\phi^{i},\cos \theta^{i})^{T}$ gives the transpose magnetic moment unit written in the $O(3)$ representation with a  unit modulus constraint $|{\bf n}^{i}|^{2} =1$. The first term in the Hamiltonian represents the exchange interaction with exchange constant $J$, the second term is the DMI term with  $D$ being  the Dzyaloshinskii-Moriya (DM) vector constant. DMI term is a manifestation of chirality in the system since it has a vanishing value for centrosymmetric crystal structures. The last term is the Zeeman interaction. According to Derrick-Hobart theorem, the Hamiltonian  \ref{effective} supports the emergence of large skyrmions \cite{Derrick,Hobart}. Suppose there exists a skyrmion solution ${\bf n}^{0}$ to the system. We compute each contribution in the energy functional  as $E^{0}_{H}$, $E^{0}_{DM}$ and $E^{0}_{Z}$, where $H, DM$ and $Z$  denote Heisenberg exchange, Dyzaloshinskii-Moriya and Zeeman terms.  Now we consider the scaling  ${\bf n}= {\bf n}^{0}(\lambda x)$. Substituting this scaled solution into each term in the energy functional gives, 
\begin{equation}\label{hobart}
E(\lambda)= E^{0}_{H}- \lambda^{-1}|E^{0}_{DM}|+ \lambda^{-2} E^{0}_{Z}.
\end{equation} 
This  has a unique minimum point which could be  found  by the relation
	$\lambda=\frac{2 E^{0}_{Z}}{|E^{0}_{DM}|}$. It is safe  to  choose $\lambda=1$ for consistency throughout our argument. From \ref{hobart}, it is not difficult to observe that  skyrmion  is    stabilized by DMI term. When $\lambda\rightarrow \infty$, the equation \ref{hobart}  implies that a skyrmion shrinks to zero without the DMI term.  The perpendicular  magnetic anisotropy (PMA) term was ignored since such a term does not play an important role in Fe$_{0.5}$Co$_{0.5}$Si \cite{Feco}. The total energy is the spatial integral of $\mathcal{H}_{i}$: $ H_{i}= \int d^{2} r\; \mathcal{H}_{i}$. 
The bilayer skyrmion can be described by the following Hamiltonian $H_{tot}=H_{S}+H_{A}+H_{int}$. The term $H_{int}$ is assumed to contain the AFM exchange coupling between the two chiral magnets, 
\begin{equation}
H_{inter}= -J_{int}\int d^{2}x\; {\bf n}^{i=S}. {\bf n}^{i=A}.
\end{equation}

AFM interaction term is responsible for the coupling between spin degrees of freedom in skyrmion and spin degrees of freedom in antiskyrmion. AFM-coupled spins are in opposite alignment with each others. \\
We will use a purely geometric approach in our investigation of quantum entanglement. Thus,  it is more convenient to work in the equivalent $\mathbb{CP}^{1}$ formulation of  nonlinear sigma model NL$\sigma$M 
\cite{Belavin, Han}. This can be achieved  using Hopf map ${\bf n}^{i}= ({\bf z}^{i})^{\dagger}{\bf \sigma}^{i} {\bf z}^{i}$. This mapping connects the  classical object ${\bf n}^{i}$ with spinor ${\bf z}^{i}=\begin{bmatrix}
\cos\frac{\theta^{i}}{2}  \\
\sin\frac{\theta^{i}}{2}e^{i\phi^{i}} 
\end{bmatrix}$. The spinor ${\bf z}^{i}$ can be interpreted as the coherent-state wavefunction of spin-$\frac{1}{2}$ particles.  The DMI term can be phrased  in term of the spinor ${\bf z}_{i}$ as follows
\begin{align}
{\bf n}^{i}.(\nabla \times {\bf n}^{i})= \sin\theta^{i} \cos\theta^{i} (\cos\phi_{i}\; \partial_{x}\phi^{i} + \sin\phi\; \partial_{y}\phi^{i})\\ \nonumber + (\sin\phi^{i} \; \partial_{x}\theta^{i} - \cos\phi^{i} \;\partial_{y}\theta^{i} -\sin^{2}\theta^{i}\partial_{z}\phi^{i})\\ \nonumber=  -2{\bf n}^{i}\cdot {\bf a}^{i} -i({\bf z}^{i})^{\dagger} (\sigma^{i}\cdot\nabla) \;{\bf z}^{i} +i(\nabla ({\bf z}^{i})^{\dagger})\cdot\; \sigma^{i} {\bf z}^{i}.
\end{align} 

The equation \ref{effective} can be re-expressed in term of the spinor ${\bf z}^{i}$ as
	\begin{align}\label{cp1}
\mathcal{H}_{i}= \frac{J}{2} \;  (\partial_{\mu}{\bf n}^{i})\cdot (\partial_{\mu}{\bf n }^{i}) + D \;{\bf n}^{i}\cdot(\nabla \times{\bf n}^{i}) -{\bf B}\cdot{\bf n}^{i}
\\ \nonumber =2 J \;  \big(\partial_{\mu} ({\bf z}^{i})^{\dagger}+ia^{i}_{\mu}\; {\bf z}^{\dagger}_{i} - i\kappa \;({\bf z}^{i})^{\dagger} \sigma^{i}_{\mu}\big) \\ \nonumber \big(\partial_{\mu} {\bf z}^{i}- ia^{i}_{\mu}\;{\bf z}^{i}+i\kappa\; \sigma^{i}_{\mu}{\bf z}^{i}\big) -{\bf B} \;  ({\bf z}^{i})^{\dagger} \sigma^{i} {\bf z}^{i} \\ \nonumber
=2J (D^{i}_{\mu}{\bf z}^{i})^{\dagger} D^{i}_{\mu}{\bf z}^{i}- {\bf B} \;  ({\bf z}^{i})^{\dagger} \sigma^{i} {\bf z}^{i}, 
\end{align}

where $D^{i}_{\mu}= \partial_{\mu}- ia^{i}_{\mu}+i\kappa\sigma^{i}_{\mu}$ denotes the covariant derivative for thin film $i$, $\kappa=\frac{D}{2J}$, and  $a^{i}_{\mu}= -i ({\bf z}^{i})^{\dagger}\partial_{\mu}{\bf z}^{i}$ is the emergent gauge field. Each magnetic layer carries a $\mathbb{CP}^{1}$- field which is responsible for the magnetization.  The $\mathbb{CP}^{1}$-field  is a two-component normalized vector with complex entries such that each field is being represented using $SU(2)$ representation.
The inclusion of DMI term in the effective Hamiltonian \ref{cp1} is done simply by adding a non-Abelian gauge field proportional to  Pauli matrices $\sigma_{\mu}$. The emergent gauge field $a^{i}_{\mu}$ is usually called the real-space Berry connection. It is synthesized by adiabatically varying the spin texture sufficiently slow in time. The real-space Berry phase connection can give rise to the skyrmion Hall effect. Unlike the momentum-space Berry connection which gives rise to the anomalous Hall effect \cite{Freimuth}.
Although the non-Abelian gauge field is non-dynamic (constant), it has an associated flux with it. The covariant derivative commutator gives the field tensor, 
\begin{equation}
F^{i}_{\mu \nu}= i[D^{i}_{\mu},D^{i}_{\nu}]=f^{i}_{\mu\nu}+2\kappa^{2}\epsilon_{\mu\nu\lambda}\; \sigma^{i}_{\lambda}.
\end{equation}
where the abelian part of the flux is  $f^{i}_{\mu\nu}=\partial_{\mu}a^{i}_{\nu}-\partial_{\nu}a^{i}_{\mu}$.

The two-dimensional emergent  vector potential for a single magnetic skyrmion is 
\begin{equation}\label{gauge}
{\bf a}^{i}= -i({\bf z}^{i})^{\dagger}\nabla_{2} {\bf z}^{i}=\frac {\hat{\phi^{i}}}{2r} (1-\cos\theta^{i}(r))= \frac{\hat{\phi^{i}}}{r} \sin^{2} \frac{\theta^{i}}{2},
\end{equation}
Since $\hat{\phi^{i}}  =(\sin\phi^{i}, \cos\phi^{i},0)$ , the gauge field ${\bf a}^{i}$ turns to be a  two-dimensional object.  The magnetic flux originating from this gauge field is 
\begin{equation}
\nabla_{2} \times {\bf a}^{i} = \frac{1}{2r} \sin\theta^{i}(r)\;(\theta^{i})^{\prime}(r),
\end{equation}
where 
 $\nabla_{2}\equiv(\partial_{x},\partial_{y},0)$.

The local spin orientation $(\theta^{i},\phi^{i})$ is related to the local coordinate system of a single skyrmion $(r,\varphi)$ such that $\theta_{i}=\theta_{i}(r)$  and $\phi_{i}=\varphi_{i}-\frac{\pi}{2}$. For seek of simplicity, we assume ${\bf B}=B\hat{z}>0$. The geometric considerations of skyrmions impose the following  boundary conditions on  $\theta_{i
}$: (a) $\theta_{i}(\infty)=0$ and (b) $\theta_{i}(0)=\pi$. The total energy of a single skyrmion (anti-skyrmion) reads \cite{Han} 
\begin{align}\label{SkyrmionE}
E_{SK}= 4\pi J \int ^{\infty}_{0} r dr \bigg[(\frac{1}{2} \frac{d\theta_{i}}{dr}+\kappa)^{2}\\ \nonumber-\kappa^{2} +\frac{\kappa}{r}\sin\theta_{i} \cos\theta_{i} +\frac{1}{4r^{2}} \sin^{2} \theta_{i}-\gamma (\cos\theta_{i}-1)\bigg], 
\end{align}  
where $\gamma=\frac{B}{2J}$. The total energy of large bilayer skyrmion is
\begin{equation}\label{tot}
E_{tot}=E_{Sk}(\theta_{S})+E_{Sk}(\theta_{A})+ E_{int}(\theta_{S},\theta_{A} ), 
\end{equation}
In $\mathbb{CP}^{1}$-formulation, the AFM interaction term takes the form  
\begin{equation}
E_{int}= -2\pi J_{int} \int _{o} ^{\infty} rdr  \cos\theta_{S}.  \cos\theta_{A}. 
\end{equation}
For sufficient AFM interaction, we have the case where  each spin in the first film is coupled with another opposite spin in the second film. This allows us to write  $\theta_{A}=\pi -\theta_{S}$ and express the total energy functional \ref{tot} in term of a single angle $\theta_{S}$ or $\theta_{A}$. 
The total energy functional \ref{tot} simplifies for fixed values of DM interaction constants $D$, exchange couplings $J$ and magnetic fields $B$ in both skyrmion and its AFM coupled antiskyrmion. It  takes the following  simple form 
\begin{align}\label{key}
E_{tot}=  4\pi J \int ^{\infty}_{0} r dr \bigg[(\frac{1}{2} \frac{d\theta^{S}}{dr})^{2}+ 2\gamma +\frac{1}{2r^{2}} \sin^{2} \theta^{S}\bigg] \\ \nonumber +2\pi J_{int}\int ^{\infty}_{0} r dr  \cos^{2}\theta^{S}  .
\end{align} 
Realistically, in order for \ref{key} to make sense is to introduce a hard cutoff $r_{Sk}$  such that $\theta_{S,A}(r)=0$ for $r\geq r_{Sk}$. Physically, $r_{Sk}$ is a half-skyrmion distance in skyrmion phase crystal or the size of skyrmion. \\

It is not difficult to show that our system has a vanishing skyrmion Hall effect. For each chiral magnet film, Thiele equation can be written  as \cite{Thiele,Ezawa}
\begin{equation}\label{thiele}
{\bf F}= {\bf G} \times ({\bf v}_{s}- \dot{{\bf R}})+ { \Gamma}_{ij}\; (\beta\; {\bf v}_{s}-\alpha\; \dot{{\bf R}}),  
\end{equation} 

where ${\bf v}_{s}$ denotes the velocity of spin polarized current, $\alpha$ is the Gilbert damping term, $\beta$ is the non-adiabatic damping constant, ${\bf R}$ represents the center of mass coordinates, ${\bf G}$ and ${ \Gamma}_{ij}$ are the Gyromagnetic vector and dissipative tensor respectively given by
\begin{eqnarray}
G_{i}= \varepsilon_{ijk}\; \int d^{2}{\bf r}\;  ({\bf n}, \partial_{i} {\bf n}, \partial_{j} {\bf n}), \\ \nonumber
\Gamma_{ij}= \int d^{2}{\bf r} \; \partial_{i}{\bf n}\; \partial_{j}{\bf n}. 
\end{eqnarray}

	We introduced the non-adiabatic spin transfer torque with parameter $\beta$ in \ref{thiele} to account for small dissipative forces that break the conservation of spin in the spin-transfer process. Since we have 
	considered an external magnetic field parallel to the $z$-direction and an
	in-plane spin polarized current, by symmetry considerations, dissipation tensor has the following simple form 
\begin{equation}
{\bf \Gamma} = \Gamma \begin{pmatrix}
1 & 0& 0 \\
0&  1 & 0 \\
0& 0 & 0
\end{pmatrix}
\end{equation}
and the Gyromagnetic vector takes the form 
\begin{equation}
{\bf G}= 4\pi Q \begin{pmatrix}
0& -1&0 \\
1& 0 & 0\\
0& 0& 0
\end{pmatrix}
\end{equation}

Note that ${\bf F}$ in equation \ref{thiele} vanishes since HDMZ action is translationally invariant ${\bf r}\rightarrow {\bf r}+\delta {\bf r}$ provided that Zeeman field is uniform. Thus, we obtain the following coupled equations 

\begin{equation}
\begin{pmatrix}
\alpha \Gamma & -4\pi Q \\
4\pi Q & \alpha\Gamma
\end{pmatrix}\begin{pmatrix}
\dot{X}\\
\dot{Y}\end{pmatrix}= \begin{pmatrix}
\beta \Gamma & -4\pi Q\\
4\pi Q & \beta \Gamma
\end{pmatrix}
\end{equation} 
This system is non-singular and always has a solution of the form 
\begin{align}
{\bf v}_{Sk}= \dot{{\bf R}}=(\dot{X},\dot{Y}) \\ \nonumber = \frac{\beta}{\alpha}{\bf v}_{s}+ \frac{\alpha-\beta}{\alpha^{3} (\frac{\Gamma}{4\pi Q^{2}})+\alpha} \big({\bf v}_{s} +  \frac{\alpha \Gamma }{4 \pi Q} \hat{z}\times {\bf v}_{s}\big). 
\end{align}
From the last result, we observe that skyrmion ( antiskyrmion) velocity ${\bf v}_{Sk}$ is a combination of drag velocity ${\bf v}_{s}$ and  Magnus term proportional  to the topological charge $Q$. Since we dealt with two thin films of chiral magnets where skyrmions in one layer has opposite charge with the second i.e. $Q_{S}=-Q_{A}$. This implies the vanishing of Magnus term contribution for the whole system.

\section{SU(4) Parametrization of Bilayer Skyrmion}\label{su4}

At the perfect coupling between skyrmion-antiskyrmion pairs such that each spin in skyrmion is AFM-coupled to a spin in antiskyrmion, the system skyrmion-antiskyrmion pair can be described by a four-component wavefunction. Thus  we  can represent the  spin degrees of freedom in bilayer skyrmions using $SU(4)$ symmetry. The $SU(4)$ skyrmions were studied before  in multicomponent quantum Hall system \cite{Tsitsishvili} and graphene \cite{Sarma}. It was found that skyrmions in these systems are stabilized mainly by the competition between Zeeman and Coulomb  interactions , unlike skyrmions in chiral magnets . However, both skyrmions share the same topological properties in common regardless of the system details. Since we have AFM coupled skyrmion-antiskyrmion pairs, our system resembles the spin-pseudospin skyrmions in term of parametrization despite the fact that we now have two skyrmions instead of one.  For large bilayer skyrmions, we consider the properties of $SU(2)\otimes SU(2)$ skyrmion-antiskyrmion pairs under the presence of DM  and Zeeman interactions (HDMZ model). We do this from a perspective of entanglement between the spin degrees of freedom in skyrmion and its AFM coupled anti-skyrmion. Because of Zeeman interaction term, the full $SU(4)$ symmetry breaks down to $U(1)\otimes U(1)$ symmetry where each symmetry group  corresponds to a rotation of spin in the skyrmion or antiskyrmion along the applied magnetic field direction ( in our case, the $z$-direction). Thus, we have the symmetry breaking sequence $SU(4)\rightarrow SU(2)\otimes SU(2) \rightarrow U(1)\otimes U(1)$.  Interestingly, the DMI term written in term of spinors ${\bf z}^{i}$ preserves the full $SU(2)\otimes SU(2)$ symmetry. This is due to the embedding of DMI term in the covariant derivative that acts on the  spinor ${\bf z}^{i}$ as a nondynamic term.

We parametrize the $SU(4)$ bilayer skyrmion using a Schmidt decomposition \cite{Goerbig}. According to Schmidt decomposition, every pure state in the Hilbert space $\mathcal{H}_{12}=\mathcal{H}_{1}\otimes \mathcal{H}_{2}$  can be written in the form 

\begin{equation}
|\psi\rangle= \Sigma_{i=0}^{N-1} \lambda_{i} |e_{i}\rangle \otimes |f_{i} \rangle ,
\end{equation} 
Where $\{|e_{i}\rangle\}^{N_{1}-1}_{i=0}$ is an orthonormal basis for $\mathcal{H}_{1}$,  $\{|f_{i}\rangle\}^{N_{2}-1}_{i=0}$ is an orthonormal basis for  $\mathcal{H}_{2}$, $N\leq min \{N_{1},N_{2}\}$, and $\lambda_{i}$ are non-negative real numbers such that $\Sigma_{i=0}^{N-1} \lambda_{i}^{2}=1$. Thus, we can express the wavefunction as 

	\begin{eqnarray}\label{wave}
	|\Psi({\bf r})\rangle= \cos\frac{\alpha}{2}\mid \varphi^{S}\rangle \otimes | \varphi^{A}\rangle+ \sin\frac{\alpha}{2} e^{i\beta} | \chi^{S}\rangle \otimes | \chi^{A}\rangle  && \\ \nonumber =\begin{pmatrix}
	\cos\frac{\alpha}{2} \cos\frac{\theta^{A}}{2} \cos\frac{\theta^{S}}{2}+ \sin\frac{\theta^{A}}{2}\sin\frac{\theta^{S}}{2} e^{i(\beta-\phi^{A}-\phi^{S})}\\
	\cos\frac{\alpha}{2}  \sin\frac{\theta^{A}}{2} \cos\frac{\theta^{S}}{2} e^{i\phi^{A}}- \sin\frac{\alpha}{2} \sin\frac{\theta^{S}}{2}\cos\frac{\theta^{A}}{2}e^{i(\beta-\phi^{S})}\\ 
	\cos\frac{\alpha}{2} \cos\frac{\theta^{A}}{2}\sin\frac{\theta^{S}}{2} e^{i\phi^{S}}- \sin\frac{\alpha}{2}\sin\frac{\theta^{A}}{2}\cos\frac{\theta^{S}}{2} e^{i(\beta-\phi^{A})} \\
	\cos\frac{\alpha}{2} \sin\frac{\theta^{A}}{2}\sin\frac{\theta^{S}}{2} e^{i(\phi^{A}+\phi^{S})}+ \sin\frac{\alpha}{2}\cos\frac{\theta^{A}}{2}\cos\frac{\theta^{S}}{2} e^{i\beta}
	\end{pmatrix} 
	\end{eqnarray}

Where $\alpha \; \epsilon\; [0,\pi]$ and $\beta  \; \epsilon\; [0,2\pi]$ are function of ${\bf r}$, 
and the local two-component spinors $|\varphi^{S}\rangle$, $| \chi^{S}\rangle$, $|\varphi^{A}\rangle$ and $|\chi^{A}\rangle$ are constructed as follows
$|\varphi^{i}\rangle=\begin{pmatrix}
\cos\frac{\theta^{i}}{2}  \\
\sin\frac{\theta^{i}}{2}e^{i\phi^{i}} 
\end{pmatrix}$ and  $|\chi^{i}\rangle =\begin{pmatrix}
-\sin\frac{\theta^{i}}{2}e^{-i\phi^{i}} \\ \cos\frac{\theta^{i}}{2}
\end{pmatrix}$, where $\theta^{i}\;\epsilon \;[0,\pi]$ and $\phi^{i}\;\epsilon \;[0,2\pi]$ are the usual polar angles defining the vector ${\bf n}^{i}$. We can read off directly the reduced density matrices using the Schmidt decomposition. The reduced density matrices for spins in skyrmion and antiskyrmion are 

\begin{eqnarray}
\rho_{S}= \mathrm{Tr}_{A}( | \Psi({\bf r})\rangle \langle  \Psi({\bf r} |) = \\ \nonumber \cos^{2}\frac{\alpha}{2} | \varphi^{S} \rangle \langle \varphi^{S}| + \sin^{2}\frac{\alpha}{2}\mid \chi^{S} \rangle \langle \chi^{S}|,
\end{eqnarray}

\begin{eqnarray}
\rho_{A}= \mathrm{Tr}_{S}( | \Psi({\bf r})\rangle \langle  \Psi({\bf r}|) = \\ \nonumber \cos^{2}\frac{\alpha}{2} \mid \varphi^{A} \rangle \langle \varphi^{A}| + \sin^{2}\frac{\alpha}{2}| \chi^{A} \rangle \langle \chi^{A}|. 
\end{eqnarray}

It is convenient to express the wavefunction \ref{wave} as $|\Psi({\bf r})\rangle= \begin{pmatrix}
	z_{1}\\ z_{2}\\ z_{3}\\ z_{4}
\end{pmatrix}$ such that entanglement measure can be written gently as \cite{Goerbig}
\begin{equation}
\mathfrak{E}= 4 |z_{1}z_{4}- z_{2}z_{3}|^{2}. 
\end{equation}
For maximal entangled states we have $z_{1}=z_{2}= \frac{1}{\sqrt{2}}$ and $z_{2}=z_{3}=0$ while for separable (factorisable) states we have $z_{1}z_{4}= z_{2}z_{3}$.  
Consider for simplicity the case when spins in skyrmion and antiskyrmion are perfectly entangled. Let $|\varphi_{S}\rangle= \begin{pmatrix}
1\\0
\end{pmatrix}$, $|\varphi_{A}\rangle= \begin{pmatrix}
0\\1
\end{pmatrix}$, $|\chi_{S}\rangle=\frac{1}{\sqrt{2}} \begin{pmatrix}
1\\1
\end{pmatrix}$ and $|\chi_{A}\rangle=\frac{1}{\sqrt{2}} \begin{pmatrix}
-1\\1
\end{pmatrix}$. Clearly when  $\alpha=\frac{\pi}{2}$, the off-diagonal terms of density matrix vanish and the diagonal terms become 1. This verify the maximal entanglement condition $\rho_{ik}^{A} =\frac{1}{2} \mathbb{I}_{2}$. \\
The local transformation operators $U$ of the density matrices form a six-dimensional subgroup $SU(2)\otimes SU(2)$  of the full unitary group $U(4)=U(1)\otimes SU(4)$.  The local transformation operators $U$ are  parametrized by an arbitrary six real variables such that $U(\theta_{S},\phi_{S},\theta_{A},\phi_{A},\alpha,\beta)^{\dagger} U(\theta_{S},\phi_{S},\theta_{A},\phi_{A},\alpha,\beta)=\mathbb{I}_{4}$ (4$\times 4$ identity matrix) . Without loss of generality, we can use $\mathbb{I}_{2} \otimes \sigma_{\mu}$ and $\sigma_{\mu}\otimes\mathbb{I}_{2}$ as hermitian $\mathfrak{su}(2)\otimes \mathfrak{su}(2)$ Lie algebra  basis of the full $SU(4)$-bilayer skyrmion theory. Here, $\sigma_{\mu}$ and  $\mathbb{I}_{2}$  denote the Pauli matrices and the two-dimensional identity matrix respectively.

\section{Geometry of quantum states } 

 We give a geometric description to the problem of entanglement in bilayer skyrmions based on complex projective geometry \cite{Bengtsson}. $\mathbb{CP}^{\mathcal{N}}$  is the space of rays in $\mathbb{C}^{\mathcal{N}+1}$, or equivalently the space of equivalence classes of $\mathcal{N}+1$ complex numbers, at least one of them is non-zero, under $(Z^{0}, Z^{1}, \dots,Z^{\mathcal{N}}) \sim \lambda (Z^{0}, Z^{1}, \dots,Z^{\mathcal{N}})$, where $\lambda \varepsilon\; \mathbb{C}$ and $\lambda \neq 0$. In quantum theory,  $\mathbb{CP}^{\mathcal{N}-1}$-field  corresponds to  $N$-component normalized spinor  $z=(z_{1},z_{2},z_{3},\dots z_{\mathcal{N}})^{T}$ such that  two vectors $z$ and $e^{i\varphi}z$ are equivalent for arbitrary $\varphi\; \epsilon \; \mathbb{R}$. The  normalization of $\mathbb{CP}^{\mathcal{N}-1}$ spinor takes away two real parameters (or one complex) which explains  why the space $\mathbb{CP}^{\mathcal{N}-1}$ correspond to $\mathbb{C}^{\mathcal{N}}$ . As we have seen in previous section, the skyrmion -antiskyrmion coupled pair can be described using a four-component spinor which lives in $\mathbb{CP}^{3}$-manifold.   Any  $\mathbb{CP}^{3}$-manifold is isomorphic to $\frac{U(4)}{[U(3)\otimes U(1)]}\cong \frac{SU(4)}{[SU(3)\otimes U(1)]}$, therefore the second homotopy group is $\pi_{2}(\mathbb{CP}^{3})= \pi_{2}\{\frac{SU(4)}{[SU(3)\otimes U(1)]}\}= \pi_{1}[SU(3)\otimes U(1)]$. Using the fact that  homotopy group for the product manifold factorizes as $\pi_{k}(\mathrm{g}\otimes \mathcal{H})= \pi_{k}(\mathrm{g})\otimes \pi_{k}(\mathcal{H})$ alongside with  the fact that any simple Lie group $\mathrm{g}$ has a vanishing fundamental homotopy group (i.e $\pi_{1}(\mathrm{g})=0$) give $\pi_{2}(\mathbb{CP}^{3})=\pi_{1}[SU(3)]\otimes \pi_{1}[U(1)]=\mathbb{Z}$ \cite{Nash, Nakahara}. \\ 
The pure state for each spin-$\frac{1}{2}$   represents a vector in the  two-dimensional complex vector space. In  Dirac notation, this vector can be expressed as $|\Psi\rangle= \Sigma _{i=0} ^{N-1}\;  Z^{i}  |i \rangle$, where $|i \rangle$ is a given orthonormal basis. The distance $D_{FS}$ between two states $|\Psi_{1}\rangle$ and $|\Psi_{2}\rangle$ is given by the Fubini-Study distance \cite{Vallee}
\begin{equation}
\cos^{2}D_{FS}=\frac{| \langle \Psi _{1}| \Psi_{2}\rangle|^{2}}{\langle \Psi_{1} | \Psi_{1}\rangle \langle \Psi_{2} | \Psi_{2}\rangle}= \frac{| Z_{1}. \overline{Z}_{2}|^{2}}{(Z_{1}. \overline{Z}_{1}) (Z_{2}. \overline{Z}_{2}) } .
\end{equation}
Where $\overline{Z}_{i}$ is the row vector whose entries are the complex conjugates of  entries in the column vector $\overline{Z}^{i}$. The Fubini-Study metric measures the distinguishability of pure quantum states. In quantum communication theory, Fubini-Study distance is known as fidelity function. Since we considered a continuum theory for describing large  bilayer skyrmions in scetion \ref{continuum}. The distinguishability of any two arbitrary states in large skyrmion or antiskyrmion is supposed to be difficult to observe.
The infinitesimal form of the Fubini-Study distance approaches  the Fubini-Study  metric tensor 
\begin{equation}
ds^{2}= \frac{Z.\overline{Z} dZ.d\overline{Z}- Z.d\overline{Z} dZ .\overline{Z}}{(Z.\overline{Z}) (Z.\overline{Z})} ,
\end{equation}
Here $Z.\overline{Z}= Z^{i}\overline{Z}_{i}$. From Fubini-Study metric, time-energy uncertainty relation can be derived directly for each single spin \cite{Bengtsson}.  
As a spin-coherent state goes through a closed loop, it will gain the phase $\gamma= \oint \langle \psi(s)\mid  \frac{d}{ds}\mid \psi(s)\rangle$. It was found that this phase is equal to the Riemannian curvature $K=\frac{1}{2S}$ of the phase space of spin-coherent state up to a constant. 
When $S=1/2$ ( like large 2D skyrmions), the curvature is equal to its maximum value $K=1$ \cite{Vallee}.\\
Any arbitrary state vector for bipartite composite system reads 
\begin{equation}
|\Psi \rangle = \frac{1}{\sqrt{N}} \Sigma _{i=0}^{N-1} \Sigma_{j=0} ^{N-1} C_{ij} | i\rangle \otimes |j\rangle ,
\end{equation}

where $C_{ij}$ is  $N\times N$ matrix with complex entries. The density matrix for the composite system is given by 
$\rho_{ij,kl}= \frac{1}{N} C_{ij}C^{\star}_{kl}$. Since the system is in pure state, its density matrix has rank one. Now suppose we perform experiment in one of the two chiral magnets, the reduced density matrix for this subsystem is the partially traced density matrix $\rho_{\mathcal{A}}=\mathrm{Tr}_{\mathcal{B}} \rho:= \mathrm{Tr}_{\mathcal{H_{B}}} \rho$  which equals to $\rho^{\mathcal{A}}_{ik}= \Sigma _{j=0}^{N-1}\rho_{ij,kj}$.  The rank of this subsystem density matrix may be greater than one. The global state of the chiral magnets system may be written as a product state spanned in the total Hilbert space $\mathcal{H}=\mathcal{H_{A}}\otimes\mathcal{H_{B}}$ 
\begin{equation}
\mid \Psi \rangle= \mid  \mathcal{A}\rangle \otimes \mid \mathcal{B}\rangle= \Sigma _{i=0}^{N-1} \Sigma _{j=0}^{N-1}  (a_{i} \mid i\rangle ) \otimes (b_{j} \mid j\rangle), 
\end{equation} 
So the matrix $C_{ij}= a_{i}b_{j}$ is the dyadic product of two vectors $a$ and $b$. It is not difficult to notice that such global state of this kind is disentangled or separable since the partially traced matrix and the matrix $C_{ij}$ have rank one and the subsystems are  in pure states of their own. On other hand, the maximally entangled state can be identified using the condition
$\rho_{ik}^{\mathcal{A}} =\frac{1}{N} \mathbb{I}_{N}$ which corresponds to $\Sigma_{j=0}^{N-1} C_{ij}C^{\star}_{kj}=\delta_{ik}$. It means that we know nothing at all about the state of the subsystems even though the global state is precisely determined. 
The maximally entangled states form an orbit of the group of local unitary transformations. In our case, this group is  $\frac{SU(2)}{\mathbb{Z}}= SO(3)$ which is identical to the real projective space $\mathbb{RP}^{3}$. In general, the group $\frac{U(\mathcal{N})}{U(1)}= \frac{SU(\mathcal{N})}{\mathbb{Z}^{\mathcal{N}}}$ is a Lagrangian submanifold of $\mathbb{CP}^{\mathcal{N}^{2}-1}$. Between these two cases, the separable and maximally entangled cases, the Von Neumann entropy $S= -\mathrm{Tr}(\rho_{\mathcal{A}} \;\mathrm{ln}\rho_{\mathcal{A}})$ takes some intermediate value with a possibility of partial entanglement between the spins.
\\

\section{Conclusion}
In this letter, the problem of antiferromagnetically coupled skyrmions (bilayer skyrmion)  has been studied in detail using continuum theory approach. This was done by considering two thin films formed from the same chiral magnet  separated by an insulating spacer with antiferromagnetic coupling between chiral  films. We assumed each chiral film to host Bloch skyrmions under certain range of temperatures and external magnetic fields determined by the film parameters.  Skyrmions in the first thin-film are equal in size with skyrmions in the second thin-film but with opposite topological charge. 
 We give a representation for the spin degrees of freedom  based on $SU(4)$ Lie algebra. Moreover, we have computed the density matrices for the spin degrees of freedom  in skyrmion and its AFM-coupled antiskyrmion using Schmidt decomposition. Utilizing from the computed density matrices, we found the  conditions for maximal or partial entanglement and separability within bilayer skyrmions .  The full  $SU(4)$  symmetry is broken to $SU(2)\otimes SU(2)$ symmetry during the Schmidt decomposition process while Zeeman interaction term causes the breaking to $U(1)\otimes U(1)$ symmetry. Interestingly DMI terms preserves the $SU(2)\otimes SU(2)$ symmetry. 
\\  The geometry of quantum states in bilayer skyrmions can be described using complex projective space $\mathbb{CP}^{3}$ endowed with a unitary-invariant Fubini-Study metric. Geometrically,  the entangled states can be described  naturally using $\mathbb{CP}^{3}$ space. We have two extreme cases corresponding to maximally entangled and separable states.  The space of   maximally entangled states  happens to be the real projective space $\mathbb{RP}^{3}$ while for separable states is simply the space $\mathbb{CP}^{1}\otimes \mathbb{CP}^{1}$. 
The entanglement in skyrmion-antiskyrmion pairs can be extended to the whole skyrmion lattice SkX. We can have maximally entangled, partially entangled and separable cases for each coupled pairs. However, for uncoupled skyrmion-antiskyrmion pairs, the formalism studied in this letter can not be accurate in term of entanglement conditions. This is mainly because of the fact that skyrmion-antiskyrmion pairs are treated as antiferromagnetically coupled pairs throughout our formalism. This allowed us to impose conditions on energy eigenstates of skyrmion and antiskyrmion accordingly.  We need to consider each skrymion and antiskyrmion as $XXZ$ spin chains and calculate the corresponding entanglement entropy\cite{Vedral}. Other  possible scenarios such as having saturated ferromagnetic phase or general helical spin phase in one layer and skyrmion lattice in the second layer are theoretically possible . However we do not find these structures to be of great interest at least in term of having a vanishing Magnus force for  the whole system.    \\In comparison with graphene and multicomponent Hall systems,  intimate relation  between the entanglement conditions in  large bilayer skyrmions and $SU(4)$-skyrmions has been found.  However the system which has been investigated in  this letter is different from that studied in Graphene and multicomponent quantum  Hall systems. For example, they dealt in graphene case with spin-valley pseudospin degrees of freedom in a single skyrmion \cite{Rosch}.  In contrast, we have considered two  skyrmions with AFM coupling between its internal spins. This is the reason why we used $\mathbb{CP}^{1}\otimes \mathbb{CP}^{1} $-theory instead of $\mathbb{CP}^{3}$-theory. However, the space of entangled states is $\mathbb{CP}^{3}$-manifold as expected \cite{Bengtsson,Brody}.

As a last comment, we propose the usage of entanglement in skyrmion-antiskyrmion lattices for probing the geometric nature of quantum entanglement. This will help in turn further understand and  possibly   manipulating  skyrmion lattices in performing quantum technological tasks such as generating entanglement in systems with large number of spins. 
\vskip 5mm
{\bf Acknowledgments}: I am very much
 grateful to Professors A.N.Bogdanov  and T.Dereli  for many fruitful 
 comments on earlier draft of this work. The author  would like to thank  the anonymous Referee for the constructive comments and for pointing out the typos. The author acknowledges Ko\c{c} University for support. 

\section{$SU(4)$ Representation} \label{basis}
The special unitary group $SU(N)$ has $(N^{2}-1)$ generators, where $-1$ is because of the condition $\mathrm{det}(M)=1$ where $M$ is any element from $SU(N)$. We denote the generators as $\lambda_{A}$, $A=1,2,\dots ,N^{2}-1$. We choose the following normalization condition between generators    $\mathrm{Tr}(\lambda_{A}\lambda_{B})= 2\delta_{AB} $. Their commutator and anti-commutator are \cite{Gell-mann}
\begin{eqnarray}
[\lambda_{A}, \lambda
_{B}]= 2i \; f_{ABC} \lambda_{C}, \\ 
\{\lambda_{A}, \lambda_{B}\}=\frac{4}{N} + 2 \; d _{ABC}\lambda_{C}, 
\end{eqnarray}
where $f_{ABC}$ and $d _{ABC}$ are the structure constants of $SU(N)$. When $\lambda_{A}= \sigma_{A}$ (Pauli matrix) we have $f_{ABC}= \varepsilon_{ABC}$ and $d_{ABC}=0$ in the case of $SU(2)$. \\
Since our developed model of bilayer skrmions in chiral  magnets is based on $SU(4)$ we will give a specific attention to this group. 
$SU(4)$ has 15 generators while $SU(2)\otimes SU(2)$ has 6 generators in total. 
Embedding $SU(2)\otimes SU(2)$ into $SU(4)$ we find the matrix representation for  skyrmion $S$ and  its AFM-coupled antiskyrmion $A$
\begin{eqnarray}
\tau_{x}^{S}= \begin{pmatrix}
\sigma_{x} & 0\\
0 & \sigma_{x}
\end{pmatrix}, \tau_{y}^{S}= \begin{pmatrix}
\sigma_{y} & 0\\
0 & \sigma_{y}
\end{pmatrix}, \tau_{z}^{S}= \begin{pmatrix}
\sigma_{z} & 0\\
0 & \sigma_{z}
\end{pmatrix},
\end{eqnarray}

\begin{eqnarray}
\tau_{x}^{A}= \begin{pmatrix}
0 & \mathbb{I}_{2}\\
\mathbb{I}_{2} & 0
\end{pmatrix}, \tau_{y}^{A}= \begin{pmatrix}
0 & -i \mathbb{I}_{2}\\
i \mathbb{I}_{2}& 0
\end{pmatrix}, \tau_{z}^{A}= \begin{pmatrix}
\mathbb{I}_{2} & 0\\
0 & -\mathbb{I}_{2}
\end{pmatrix}.  
\end{eqnarray}


\begin{thebibliography}{99}


\bibitem{Nagaosa}
\Name{Nagaosa N and Tokura Y} 
\REVIEW{Nat. Nanotech.}{8}{2013}{899-911}. 

\bibitem{Nash}
\Name {Nash C and  Sen S} 
``Topology and Geometry for Physicists'', Academic Press 1983.


\bibitem{Skyrme1}
\Name{ Skyrme T.H.R} 
\REVIEW{Proc. Roy. Soc. Lond. A}{260}{1961}{127-138}.

%%%%%%%


\bibitem{Muhlbauer}
\Name{M{\"u}hlbauer S. et al.}
\REVIEW{Science}{323}{2009}{915-919 }.

\bibitem{Bogdanov}
\Name{Bogdanov A. N.  and  Yablonskii D. A. } 
\REVIEW{Sov.Phys.JETP}{68}{1989}{101-103}.


\bibitem{Bogdanov1}
	\Name{ Bogdanov A. N. and  Hubert A.}
	\REVIEW{J. Magn. Magn. Mater.}{138}{1994}{255-269}.

\bibitem{torque}
\Name{ Jonietz F. et al.}
\REVIEW{ Science}{330 }{2010}{1648-1651}.

\bibitem{SkHE}
	\Name{Zang J., Mostovoy M.,  Han J. H. and  Nagaosa N.} 
	\REVIEW{Phys. Rev. Lett.}{107}{2011}{136804}.

\bibitem{SkHE1}
	\Name{ Jiang W. et al.}
	\REVIEW{Nat.Phys.}{13}{2017}{162 }.

\bibitem{Sonin}
	\Name{ Sonin E.B.}
	\REVIEW{Phys. Rev. B}{55}{1997}{485}.

\bibitem{Freimuth}
	\Name{ Freimuth F., Bamler R. ,  Mokrousov Y. and Rosch A. }
	\REVIEW{ Phys.Rev.B}{88}{2013}{214409}. 

\bibitem{Ezawa}
	\Name{Zhang X. , Zhou Y. and  Ezawa M.} 
	\REVIEW{Nat.Comm.}{7}{2016}{10293}.

\bibitem{Takashima}
	\Name{Takashima R. , Ishizuka H. and Balents L.} 
	\REVIEW{Phys. Rev. B}{94}{2016}{134415}.
\bibitem{Gilbert}
 Gilbert T.L.  
\REVIEW{Phys. Rev.}{100}{1955} {1243}.

\bibitem{Thiele}
	\Name{ Thiele A. A.} 
	\REVIEW{Phys. Rev. Lett.}{ 30}{1973}{230}.

\bibitem{Feco}
	\Name{ Yu X.Z. , Onose Y. , Kanazawa N. ,  Park J. H.,  Han J. H. ,  Matsui Y.,  Nagaosa N., and Tokura Y. }
	\REVIEW{Nature}{465}{2010}{901}.


%%%%%%%%%%

\bibitem{Derrick}
	\Name{ Derrick G. H.}
	\REVIEW{J.
		Math. Phys.}{5}{1964}{1252}.

\bibitem{Hobart}
	\Name{ Hobart R. H.} 
\REVIEW{Proc. Phys. Soc. Lond.}{82}{1963}{201}.


\bibitem{Belavin}  
	\Name{Belavin A. A. , Polyakov  A. M. }
	\REVIEW{JETP Let.}{22}{1975}{10}.


 
\bibitem{Han}
	\Name{ Han J. H.,  Zang J.,  Yang Zhihua,  Park J., and  Nagaosa N.}
	\REVIEW{Phys. Rev. B}{ 82}{2010}{094429}. 
 
\bibitem{Tsitsishvili}
	\Name{Ezawa Z. F.  and Tsitsishvili G. } 
	\REVIEW{ Phys.Rev. B}{70}{2004}{125304}.
 
%%%%%%


\bibitem{Sarma}
	\Name{Yang K.,  Das Sarma S., and   MacDonald A. H.}
	\REVIEW{Phys. Rev. B}{74}{2006}{075423}. 

\bibitem{Goerbig}
	\Name{Doucot B. ,  Goerbig M. O.,  Lederer P. and  Moessner R.}  
	\REVIEW{Phys. Rev. B}{78 }{2008}{195327}.

\bibitem{Bengtsson}
	\Name{ Bengtsson I. and  {\.Z}yczkowski K.}
	``Geometry of Quantum States: An Introduction to Quantum Entanglement'', Cambridge University Press  2017.
\bibitem{Nakahara}
\Name{Nakahara M.} `` Geometry, topology, and physics'', CRC Press, Second  Edition 2003. 
\bibitem{Vallee}
	\Name{ Provost J. P. and  Vallee G.}
	\REVIEW{Comm. Math. Phys.}{76}{1980}{289-301}.

\bibitem{Rosch}
	\Name{ Lian Y.,  Rosch A., and  Goerbig M. O.}
	\REVIEW{ Phys. Rev. Lett.}{117}{2016}{056806}.
		\bibitem{Vedral}
	\Name{ Amico L. ,  Fazio R.,  Osterloh A., and Vedral V.}   
	\REVIEW{Rev.Mod.Phys.} { 80} {2008},{517-576} .
\bibitem{Brody}
\Name{ Brody D.C. and   Hughston L. P.} \REVIEW{J. Geom. Phys.} { 38} {2001}  {19-53}.
\bibitem{Gell-mann}
	\Name{Gell-Mann M. and  Ne’eman Y.}  ``The Eight-Fold Way'', Benjamin
New York 1964.





\end{thebibliography}
\end{document}